\documentclass[prl,aps,twocolumn,amsfonts,showpacs,superscriptaddress,floatfix]{revtex4}

\usepackage{epsfig}
\usepackage{psfrag}
\usepackage{color}
\usepackage{graphicx}
\usepackage{subfigure}

\begin{document}


\title{Predicting Excitonic Gaps of Semiconducting Single Walled Carbon Nanotubes From a Field Theoretic Analysis}
\author{Robert M. Konik}
\affiliation{Condensed Matter Physics and Material Science Department, Brookhaven National Laboratory, Upton,
NY 11973}
\author{Matthew Y. Sfeir} 
\affiliation{Center for Functional Nanomaterials, Brookhaven National Laboratory, Upton, NY 11973}
\author{James A. Misewich}
\affiliation{Condensed Matter Physics and Material Science Department, Brookhaven National Laboratory, Upton,
NY 11973}

\begin{abstract}
We demonstrate that a non-perturbative framework for the treatment of the excitations of single walled carbon nanotubes
based upon a field theoretic reduction is able to accurately describe experiment observations of the absolute values of excitonic energies.  
This theoretical framework yields a simple scaling function from which the excitonic
energies can be read off.  This scaling function is primarily determined 
by a single parameter, the charge Luttinger parameter of the tube, which
is in turn a function of the tube chirality, dielectric environment,
and the tube's dimensions, thus expressing disparate influences on the excitonic energies in a unified fashion. 
We test this theory explicitly on the data reported in NanoLetters {\bf 5}, 2314 (2005) and Phys. Rev. B. {\bf 82}, 195424 (2010)
and so demonstrate the method works over a wide range of reported excitonic spectra.
\end{abstract}

\maketitle

\newcommand{\del}{\partial}
\newcommand{\ep}{\epsilon}
\newcommand{\clsd}{c_{l\sig}^\dagger}
\newcommand{\cls}{c_{l\sig}}
\newcommand{\cesd}{c_{e\sig}^\dagger}
\newcommand{\ces}{c_{e\sig}}
\newcommand{\up}{\uparrow}
\newcommand{\down}{\downarrow}
\newcommand{\il}{\int^{\tilde{Q}}_Q d\la~}
\newcommand{\ilp}{\int^{\tilde{Q}}_Q d\la '}
\newcommand{\ik}{\int^{B}_{-D} dk~}
\newcommand{\ila}{\int d\la~}
\newcommand{\ilpa}{\int d\la '}
\newcommand{\ika}{\int dk~}
\newcommand{\tQ}{\tilde{Q}}
\newcommand{\rh}{\rho_{\rm bulk}}
\newcommand{\ri}{\rho^{\rm imp}}
\newcommand{\sh}{\sig_{\rm bulk}}
\newcommand{\si}{\sig^{\rm imp}}
\newcommand{\rph}{\rho_{p/h}}
\newcommand{\sph}{\sig_{p/h}}
\newcommand{\rp}{\rho_{p}}
\newcommand{\sip}{\sig_{p}}
\newcommand{\drph}{\delta\rho_{p/h}}
\newcommand{\dsph}{\delta\sig_{p/h}}
\newcommand{\drp}{\delta\rho_{p}}
\newcommand{\dsp}{\delta\sig_{p}}
\newcommand{\drh}{\delta\rho_{h}}
\newcommand{\dsh}{\delta\sig_{h}}
\newcommand{\enp}{\ep^+}
\newcommand{\enm}{\ep^-}
\newcommand{\enpm}{\ep^\pm}
\newcommand{\enph}{\ep^+_{\rm bulk}}
\newcommand{\enmh}{\ep^-_{\rm bulk}}
\newcommand{\enpi}{\ep^+_{\rm imp}}
\newcommand{\enmi}{\ep^-_{\rm imp}}
\newcommand{\enh}{\ep_{\rm bulk}}
\newcommand{\eni}{\ep_{\rm imp}}
\newcommand{\sig}{\sigma}
\newcommand{\la}{\lambda}
\newcommand{\ua}{\uparrow}
\newcommand{\da}{\downarrow}
\newcommand{\ed}{\epsilon_d}

One of the most challenging problems in studying low dimensional strongly correlated systems is the quantitative prediction
of the absolute values of the energies of its fundamental excitations.  These energies are typically non-perturbative in nature and so lie out of the reach of approximations that treat interactions as weak.  One non-perturbative theoretical tool that is not so limited
is quantum field theory.  Quantum field theories arise as descriptions of condensed matter systems by focusing on their low energy
properties. They have had considerable success in studying a number of problems in quantum magnetism \cite{Dender, Oshi,Essler,Essler1,Essler2,spinladders,Oshi1}, in particular the remarkable prediction of an $E_8$ symmetry in a critical quantum Ising model in a longitudinal field \cite{Zamo} that has been recently observed \cite{Coldea}, 
one dimensional Mott insulator physics \cite{Tsvelik,Tsvelik1,SO(8)ladder,Hubbard}, 
and Luttinger liquids in all of their various forms \cite{quantumwires,egger,kane1,smitha,helicaledgestates,luttingerjunction,levtsv}.
However quantum field theories are best at predicting universal properties of
materials.  Typically they do not attempt to understand absolute values of gap energies, but instead are satisfied with (the
still very non-trivial task of) computing ratios of excitation energies.

In this letter we show that this restriction need not always hold.  We demonstrate that the data that can be extracted from a field
theoretic analysis can in fact be used to predict the absolute magnitude of excitation gaps.  To this end we analyze a field theoretic treatment of the excitonic spectra of semi-conducting carbon nanotubes \cite{konik}.   The excitonic gaps of semiconducting carbon nanotubes are known to be both variegated, depending on tube diameter, chirality, subband, and dielectric environment \cite{sfeir_nano,bnl_nano,wang,Bachilo,3+4,flu,rmp}.  They are also known
to be strongly renormalized by Coulomb interactions from their bare, non-interacting values \cite{ando,kane,wang,louie,vasily}.  Both of these features make them an ideal testing ground for the analysis presented herein.

Typically excitonic spectra of carbon nanotubes have been determined using a Bethe-Salpeter equation combined with first
principle input \cite{louie,vasily,vasily1,molinari,louie1}.  While this methodology results in an estimate for the absolute magnitude of an excitonic gap, it does so by focusing upon a particular subband of a tube of a particular chirality and in a particular dielectric environment.  In our field theoretic treatment of excitonic spectra, even though we are interested in the absolute values of gaps, we are still able
to derive a universal scaling function from which the values of the excitonic gaps can be read off.  The key parameter of this
scaling function will be the total charge Luttinger parameter, $K_{c+}$, a measure of the effective strength of Coulomb interactions in the tube \cite{egger,kane}.

We begin our field theoretical treatment by focusing on a single subband of a carbon nanotube.  It is straightforward to argue that
intersubband interactions lead only to very weak perturbations on the spectra of a single subband \cite{epaps}.
To describe this subband at low energies we introduce four
sets (two for the spin, $\sigma$, degeneracy and two for the valley, $\alpha=K,K'$, degeneracy)
of right ($r=+$) and left ($r=-$) moving fermions, $\psi_{r\alpha\sigma}$.  The Hamiltonian
governing these fermions can be written as $H = \int dx ({\cal H}_{kin} + {\cal H}_{gap}) + H_{Coulomb}$.
${\cal H}_{kin}+{\cal H}_{gap}$ together give the non-interacting 
band dispersion, $\epsilon^2(p)=v_0^2p^2+\Delta_0^2$:
\begin{eqnarray}
{\cal H}_{kin} = -iv_0\psi^\dagger_{r\alpha\sigma}\partial_x\psi_{r\alpha\sigma};~~
{\cal H}_{gap} = \Delta_0\psi^\dagger_{r\alpha\sigma}\psi_{-r\alpha\sigma},
\end{eqnarray}
where $v_0$ is the bare velocity of the fermions 
and repeated indices are summed.
For the Coulombic part of the Hamiltonian we only consider the strongest part of
the forward scattering term:
$$
H_{Coulomb} = \frac{1}{2}\int dx dx' \rho(x)V_0(x-x')\rho(x'),
$$
where $\rho(x) = \sum_{r\alpha\sigma}\psi^\dagger_{r\alpha\sigma}(x)\psi_{r\alpha\sigma}(x)$.
The remaining Coulombic terms only affect the excitonic gaps at the $1\%$ level.

A unique feature of our theoretical approach is that we take into account the Coulomb interactions
at the start of the analysis.
In particular, instead of treating $H_{Coulomb}$ as a perturbation of
$\int dx ({\cal H}_{kin}+{\cal H}_{gap})$, we treat $\int dx {\cal H}_{gap}$ 
as a perturbing term of $\int dx {\cal H}_{kin}+H_{Coulomb}$.  The ``unperturbed''
Hamiltonian is 
nothing more than the Hamiltonian of a metallic carbon nanotube while
${\cal H}_{gap}$ is treated as a confining interaction on top of the metallic tube.  
To proceed in this fashion works because we can treat $\int dx {\cal H}_{kin}+H_{Coulomb}$
exactly using bosonization.

If we bosonize $H_0$ in terms of chiral bosons $\phi_{r\alpha\sigma}$
by writing $\psi_{r\alpha\sigma}\sim \exp(i\phi_{r\alpha\sigma})$, 
we arrive at a simple result \cite{kane1,egger}.  The theory
is equivalent to four Luttinger liquids described by the four bosons $\theta_{i}$, $i=c_{\pm},s_{\pm}$
(and their duals $\phi_i$)
\begin{equation}
H_{0} = \int dx\sum_{i}\frac{v_i}{2}\bigg(K_i(\partial_x\phi_{i})^2+K_i^{-1}(\partial_x\theta_i)^2\bigg).
\end{equation}
The four bosons diagonalizing $H_0$ are linear combinations of the original four
bosons and represent an effective charge-flavour separation
where $\theta_{c+} = \sum_{r\alpha\sigma}r\tilde\psi_{r\alpha\sigma}$
is the charge boson and the remaining three bosons reflect the spin, valley, and parity symmetries
in the problem.  The charge boson is the only boson to see the effects of the Coulomb interaction.
Both the charge Luttinger parameter, $K_{c+}$, and the charge velocity, $v_{c+}=v_0/K_{c+}$, are strongly renormalized.
We make note here that $K_{c+}$ is the key parameter and will be the one by which we organize all of our results.

For long range Coulomb interactions, $K_{c+}$ takes the form
\begin{equation}
K_{c+} = \bigg(1+\frac{8e^2}{\pi\kappa\hbar v_0}\big(-\log(k_{\rm min}R) + c_0\big)\bigg)^{-1/2},
\end{equation}
where $\kappa$ is the dielectric constant of the medium surrounding the tube, $k_{\rm min}$ is minimum allowed wavevector in the tube,
$R$ is the tube's radius, and $c_0$ is a wrapping vector dependent $O(1)$ constant \cite{smitha,egger}.
$k_{\rm min}$ necessarily has to be larger than $2\pi/L$ where $L$ is the length of the tube, but can in principle
be much larger, say on the order of the inverse mean free path in the tube.
In typical nanotubes $K_{c+}$ can take on values in the range of $\sim .2$.  
The remaining Luttinger parameters, $K_i$, $i=c_-,s_\pm$ retain their non-interacting values, 
$1$, and 
so their velocities, $v_i=v_0$ go unrenormalized. 
\begin{figure}[t]
\centering
\includegraphics[width=0.5\textwidth]{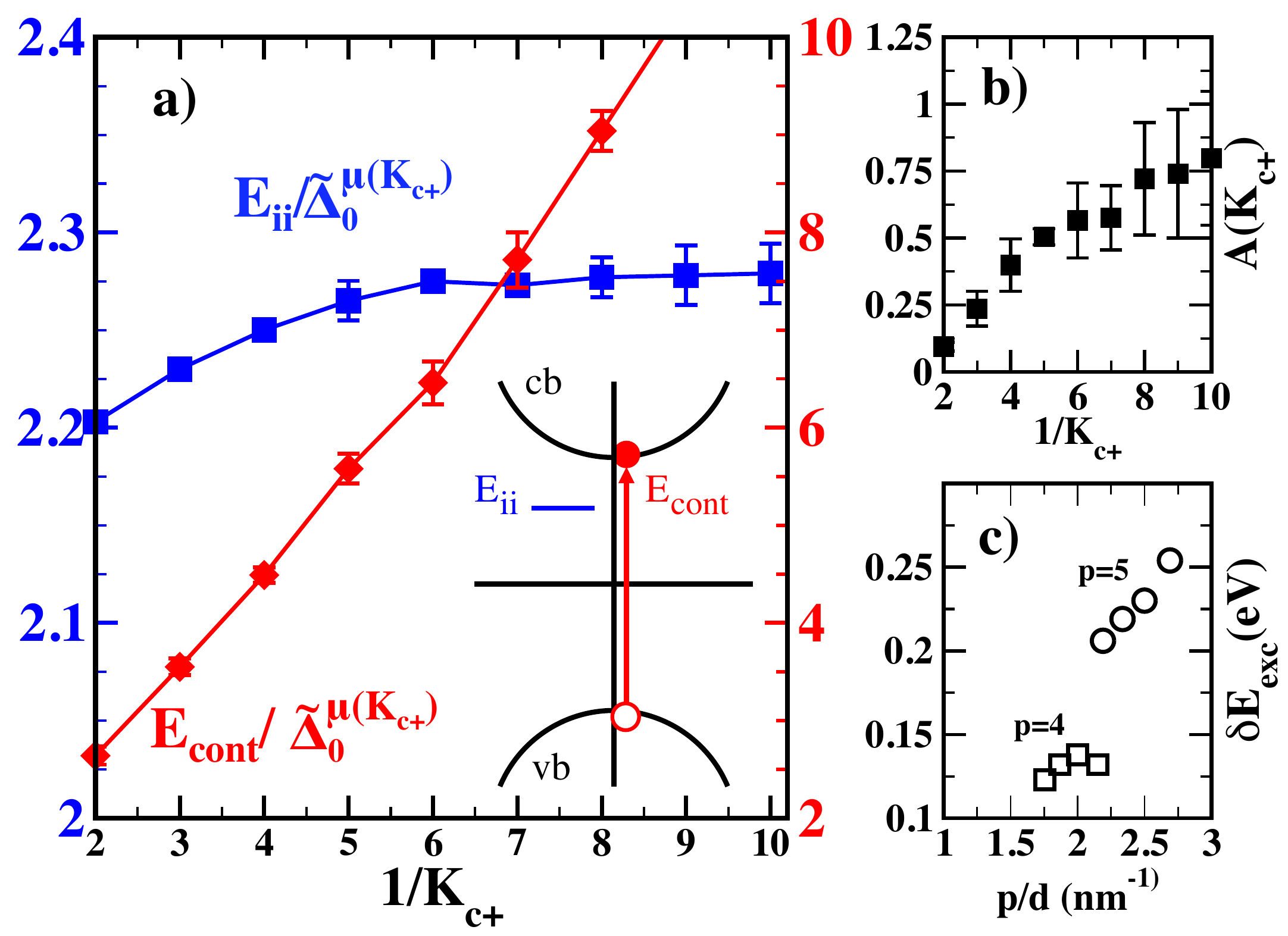}
\caption{a) The scaling functions (see Eqns. 6 and 7) for the $E_{ii}$ excitons and the particle-hole
continuum, $E_{cont}$.  At $K_{c+}=1$ (the non-interacting limit)
these functions go to $2$. Inset to a) Sketch of $E_{ii}$ and $E_{cont}$ excitations.  b) The function
$A(K_{c+})$ giving the size of the finite bandwidth correction to $E_{ii}$.  c) The size of this
correction, $\delta E_{exc}$, for the excitons, $E_{44}$ and $E_{55}$, of the four tubes studied in Ref.\cite{bnl_nano}.}
\label{scalingfunction}
\end{figure}

Under bosonization ${\cal H}_{gap}$ becomes
\begin{equation}\label{hgap}
{\cal H}_{gap} = \frac{4\tilde\Delta_0}{\pi}(\prod_{i}\cos(\frac{\theta_i}{2}) + \prod_i\sin(\frac{\theta_i}{2})),
\end{equation}
where $\tilde\Delta_0=\Delta_0(\Lambda/v_{c+})^{(1-K_{c+})/4}$ and $\Lambda$ is the bandwidth
of the tube.  This renormalization of the bandwidth has important consequences for the excitonic physics of the
tube.  In field theoretic language, the coupling $\Delta_0$, has picked up an anomalous dimension.  Rather than purely having
the dimensions of energy, $\tilde\Delta_0$ now has the dimensions of 
${\rm energy}^{(5-K_{c+})/4}\times {\rm velocity}^{(K_{c+}-1)/4}$.  This means
that all excitations gaps of the tube no longer linearly scale with $\tilde\Delta_0$ but scale rather with
the non-trivial power $\tilde\Delta_0^{4/(5-K_{c+})}$.  Coupling constants (here the bare gap) inheriting ``anomalous dimensions'' is
a standard feature of quantum field theories.  These anomalies allow one to easily access aspects of non-perturbative
physics: an immediate consequence of this was argued in Ref. \cite{konik} 
to be that the ratio of excitons between the first and second subbands goes as $2^{4/(5-K_{c+})}$ (not 2 as predicted
by non-interacting band theory), so providing a straightforward resolution of what 
Ref. \cite{kane} termed the exciton ratio problem.

While this explanation of the ratio problem required no information of how the cutoff, $\Lambda$, depends on the tube
parameters, we need more to be able to make quantitative predictions of the exciton gaps in any given
tube.  $\Lambda$ reflects the largest energy scale in the low energy reduction of the tube.  This energy
scale is not the bandwidth of graphene ($\sim 9$eV), but rather some much smaller scale reflecting that the
electrons on the tubes are delocalized around the tube's circumference.  We thus take as an ansatz
\begin{equation}\label{cutoff}
\frac{\Lambda}{v_{c+}} = \frac{B}{d},
\end{equation}
where $B$ is an $O(1)$ dimensionless constant and $d=2R$ is the tube's diameter.  We cannot directly determine this constant, but treat it
as a fitting parameter, the only undetermined parameter of this approach.  
But we will see the same constant works over tubes with a wide variety of radii, different
subbands within the same tube, and tubes in different dielectric environments.  We will also see, as an important
self-consistency check, that this same relation determines the finite bandwidth corrections to the excitonic gaps
of higher subbands.  

To this end the gap, $E_{\alpha}$, of any excitation $\alpha$ (exciton, single particle, or otherwise) takes the following 
universal scaling form
\begin{equation}
E_{\alpha} = f^\Lambda_{\alpha}(K_{c+}) \tilde\Delta_{0}^{4/(5-K_{c+})}v_{c+}^{\mu(K_{c+})},
\end{equation}
with $\mu(K_{c+})=(1-K_{c+})/(5-K_{c+})$ and where
$f^\Lambda_{\alpha}(K_{c+})$ is a scaling function.  It takes the form
\begin{equation}\label{scaling}
f^\Lambda_{\alpha}(K_{c+}) \!=\! f^\infty_\alpha(K_{c+})\big(1\!+\!A(K_{c+})(\frac{\tilde\Delta_0}{v_{0}})^{2}(\frac{v_0}{\Lambda})^{\frac{5-K_{c+}}{2}}\big).
\end{equation}
$f^\infty_\alpha(K_{c+})$ governs the gap in the large bandwidth, $\Lambda \gg \Delta_0$, limit
and was already determined in \cite{konik}.  However not previously considered, the scaling function sees corrections at finite
bandwidth.  These corrections
will be important for predicting accurately the excitonic gaps of excitons in the higher subbands.
The constant $A(K_{c+})$ is a dimensionless parameter (but depends on the charge Luttinger parameter) that governs
the size of these corrections.  It is plotted in Fig. 1b.  The form of $A(K_{c+})$ is derived in Ref. \cite{epaps}.

To extract the scaling function $f^\Lambda(K_{c+})$, we numerically study the full Hamiltonian, ${\cal H}_0 + {\cal H}_{gap}$,
using a truncated conformal spectrum approach (TCSA) \cite{zamo}
combined with a Wilsonian renormalization group
\cite{kon1}.   The results for the scaling functions for the optically active excitons, $E_{ii}$, and the particle-hole
continuum, $E_{cont}$, are shown in Fig. 1a as a function of $K_{c+}$.  We see the scaling function for $E_{ii}$ is relatively
flat as a function of $K_{c+}$ while that of $E_{cont}$ varies comparatively sharply.  
In the limit $K_{c+}$ tends to 1 (the non-interacting limit), the scaling function $f^\infty_{ii}$ tends
to 2 (i.e. the exciton energy is that of the bare (non-interacting) particle-hole continuum gap).  In the limit
$K_{c+}$ tends to 0, $f^\infty_{ii} \propto K_{c+}^{-1/5}$, only going to infinity slowly.  In contrast,
$f^\infty_{cont}$, grows much more quickly, going as $K_{c+}^{-1}$.  We have
thus quantified the general observation \cite{kane} that the renormalization of the single particle gap due to Coulomb interactions is much more marked than that of the excitons.  It is also much stronger than has been suggested in RPA-type computations
\cite{Ando}.  We also immediately infer that the binding energy of the exciton as a fraction of
the exciton energy grows as $K_{c+}^{-4/5}$ as $K_{c+}\rightarrow 0$.

\begin{figure}[t]
\center{\includegraphics[width=0.45\textwidth]{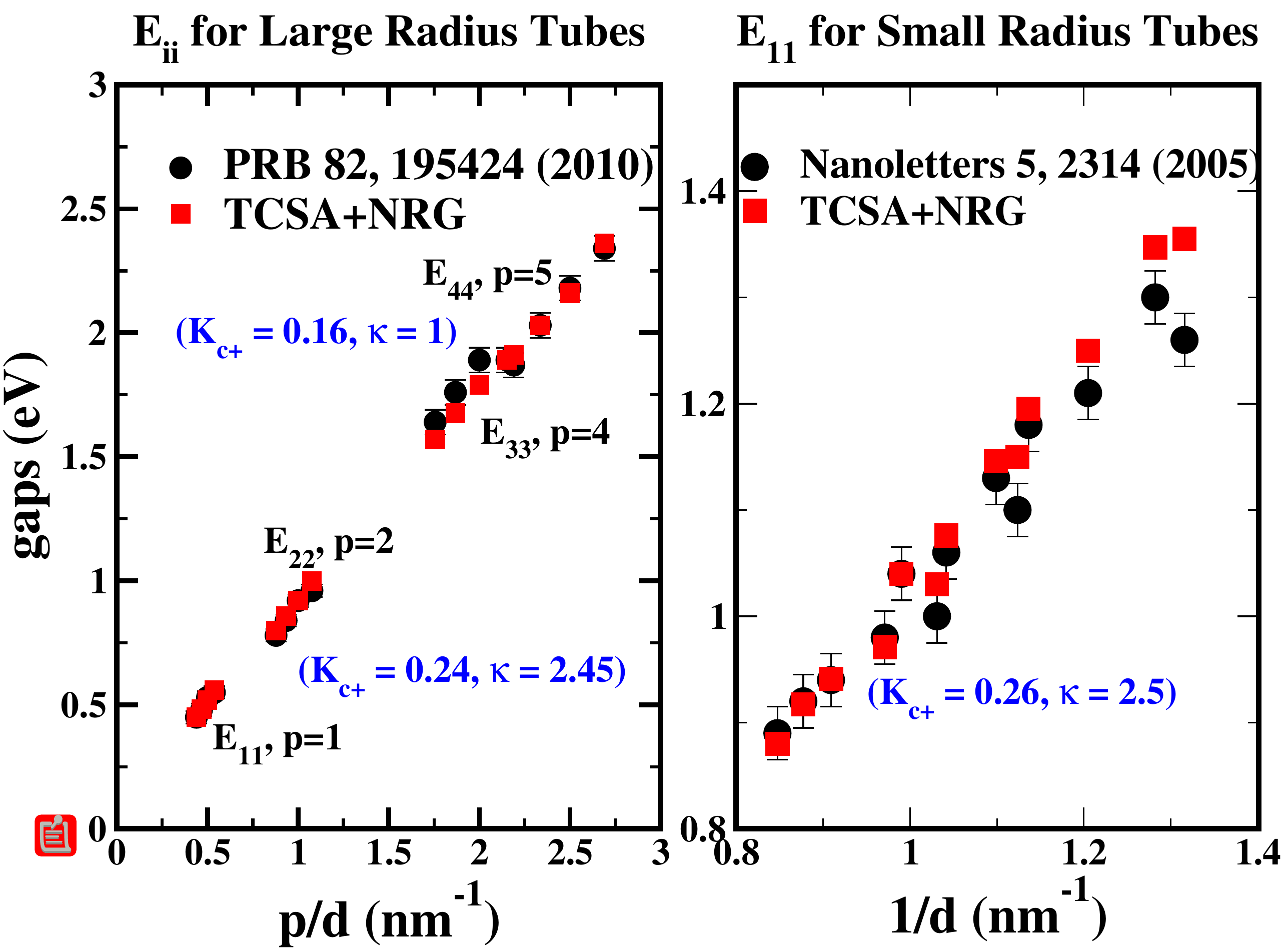}}
\caption{Left: Comparison of the measured exciton gaps of the first four subbands, $E_{ii}$, $i=1,2,3,4$
($p=1,2,4,5$ in the notation of Ref. \cite{kane}),
reported in Ref. \cite{bnl_nano}
of four nanotubes with different chiralities with gaps derived
from the scaling function determined by NRG+TCSA.  Right: Same
comparison but the measured excitonic gaps are of the first subband,
$E_{11}$, in a set of small radius tubes as reported in Ref. \cite{sfeir_nano}.}
\label{th_vs_exp}
\end{figure}

{\noindent \bf Analysis of Experimental Data}: We now examine how this theoretical approach fares in predicting
the excitonic data of Refs. \cite{bnl_nano} and \cite{sfeir_nano}.
These papers present excitonic gaps of tubes for a wide range of diameters and subbands as well as different dielectric
environments.  This will allow us to test the flexibility of the above theoretical scheme.

In Ref. \cite{bnl_nano} measurements were performed on a set of four larger diameter tubes (d
running from 1.86nm to 2.14nm).  In each of the four tubes, the first four single photon excitons, $E_{ii}$, $i=1,\cdots,4$,
were measured.  $E_{33}$ and $E_{44}$ were studied by suspending the nanotubes and using Rayleigh scattering spectroscopy.
In these measurements the relevant dielectric constant was $\kappa=1$.  These same tubes were then printed
onto a silicon wafer where source and drain electrons were patterned.
This enabled $E_{11}$ and $E_{22}$ to be measured
by a complimentary technique, Fourier-transform photoconductivity.  In this configuration the effective dielectric constant of the tubes
is the average of air and silicon dioxide, $\kappa = (1+\kappa_{\rm SiO_2})/2 = 2.45$, a result easily derived from considering
the effective potential between two charges confined to the interface of two media with different dielectric constants.
To determine the appropriate value of the Luttinger parameter for these tubes, we need to specify $k_{min}$.  
The length, $L$, of the tubes of Ref. \cite{bnl_nano} was typically $L\sim 2\mu$ (a number equal to the mean free path \cite{mft}) and so we
take $k_{min}=2\pi/l$.  This leads to a Luttinger parameter
of $K_{c+}=0.16$ for the tubes suspended in air and $K_{c+}=0.24$ for the tubes printed onto the silicon substrate.   The smaller
$K_{c+}$ for the suspended tubes indicates the action of a considerably stronger effective Coulomb interactions
for the tubes in this configuration.

In Ref. \cite{sfeir_nano} the single photon excitons, $E_{11}$, were measured for a set of 13 tubes with diameters between 0.78nm
and 1.18nm.  The tubes were embedded in a polymaleic acid/octyl vinyl ether (PMAOVE) matrix with an effective dielectric
constant of $\kappa$=2.5 \cite{wang}.  The excitons were measured using two photon spectroscopy
-- thus the $E_{2g}$ photons were also studied in this work but will not be considered here.  
As the length of tubes in the PMAOVE matrix was reported to be $L=400$nm \cite{columbia},
far smaller than $l_{mf}$, we take $k_{min}=2\pi/L$ here.  This leads to a Luttinger parameter
of $K_{c+} = 0.26$.  Because of the logarithmic
dependence of $K_{c+}$ upon $k_{min}$, $K_{c+}$ is relatively insensitive to ${\cal O}(1)$ changes of $k_{min}$.

As an ingredient to our analysis of the data in Refs. \cite{bnl_nano,sfeir_nano} we need to determine
the bare value of the gap, $\Delta_0$, for each tube.  We do so with a tight binding model based on wrapping a honeycomb
lattice of nearest neighbor spacing $a_0=1.42$A$^o$ and hopping parameter $t=3.0$eV.
We do not attempt to include curvature, twist, or stress corrections to $\Delta_0$ (\cite{eggert,yang}) although
for small radius tubes such corrections may not be insignificant.  But to be able to do so would require detailed
characterization of the tubes in their environment which is not available.
As we have explained the treatment has one fitting parameter: the constant $B$ governing the 
relationship between the effective bandwidth of the nanotube and the tube's diameter, $d$.  To find this constant $B$
we focus on the four $E_{11}$ excitonic gaps reported in \cite{bnl_nano}.  We focus on these gaps because for these
the correction due to finite bandwidth (the second term in Eqn. \ref{scaling}) can safely be ignored.  
When we fit  Eqn. (\ref{cutoff}) we find
$B\sim 0.51$.  We will henceforth use this value for B to determine theoretical values of 
the gaps for all the other
single photon excitons reported in Refs. \cite{bnl_nano,sfeir_nano}.

Remarkably this relationship between the bandwidth and the tube's diameter leads to excellent values for the other
excitonic gaps considered in this study.  To demonstrate this we first consider all (16) of the gaps reported in Ref. \cite{bnl_nano}.
Our results for the gaps are presented in Table 1 of Ref. \cite{epaps} and Fig. 2a.  We see that the agreement between the theoretically predicted
values of the gaps and the corresponding experimentally measured values is less than $2\%$ for
the $E_{11}$, $E_{22}$, $E_{44}$, and one of the $E_{33}$ gaps and on the order of $5\%$ for the remaining $E_{33}$ gaps.   
The relatively good agreement found for the $E_{33}$ and $E_{44}$
gaps is a result of taking into account the finite bandwidth corrections coming from the subleading
term in Eqn. (\ref{scaling}) where the corrections for $E_{33}$ and 
$E_{44}$ are large ($> 100$ meV -- see Fig. 1c).  These corrections in Eqn. (\ref{scaling}), inasmuch as they are proportional to
$\tilde \Delta_0^2/\Lambda^{(5-K_{c+})/2}$, depend in turn upon our identification of $\Lambda$ with
the tube diameter.  It is an important consistency check for this ansatz in Eqn. (\ref{cutoff}) 
that the computed corrections lead to a good match between the experiment and theory. 

\begin{figure}[t]
\center{\includegraphics[width=0.45\textwidth]{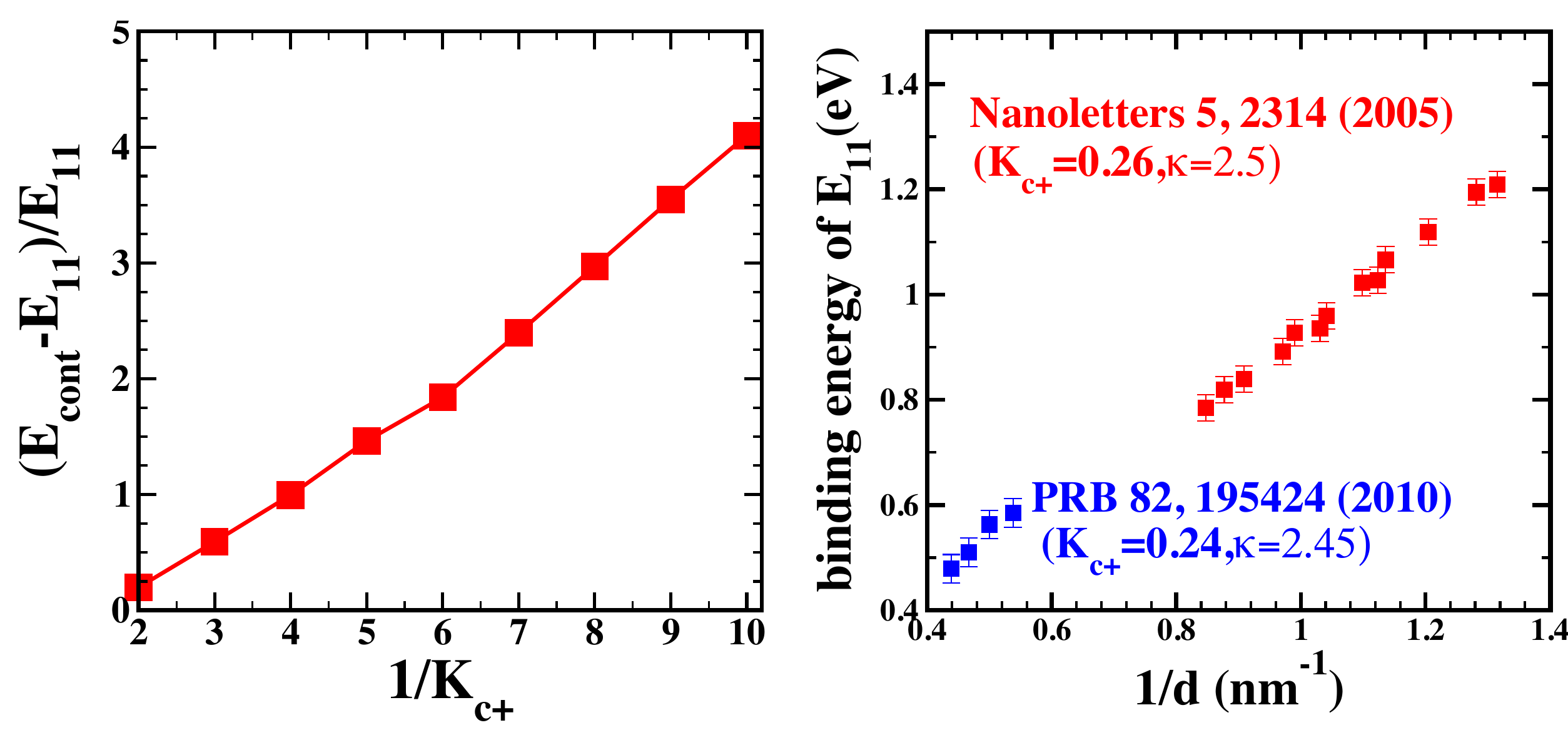}}
\caption{Left: Binding energy of exciton expressed in units of the excitonic gap.
Right: The predicted binding energies of the excitonic gaps, $E_{11}$, reported in Refs. \cite{bnl_nano,sfeir_nano}.}
\label{fig:setup}
\end{figure}

We find similar good agreement in our theoretical analysis of the $E_{11}$ excitons reported in
Ref. \cite{bnl_nano}.  Using the same relationship of
the bandwidth, we plot our predicted values for $E_{11}$ against those measured in Ref. \cite{sfeir_nano} in Fig. 3.  We see
that only for the three smallest radius tubes do we not obtain excellent agreement between
theory and experiment (as explained earlier a likely consequence of missing curvature, strain, and twist effects on
the bare gap, $\Delta_0$).  It is important to stress that the same value of $B$ derived from the four $E_{11}$ large
radius tube gaps in Ref. \cite{bnl_nano} leads to an accurate prediction of the gaps for the smaller radius tubes in Ref. \cite{sfeir_nano}.

{\noindent \bf Excitonic Binding Energies:}  We finally consider the excitonic binding energies of the $E_{11}$ excitons reported
in Refs. \cite{bnl_nano,sfeir_nano}.  
We first plot in Fig. 3a the excitonic binding energies as a function of $K_{c+}^{-1}$.  The binding energies are presented as a fraction
of the exciton gap.  We see that for $K_{c+}^{-1}$ large, the binding energies can be many multiples of the excitonic gap itself.
As $K_{c+}^{-1}$ decreases, the fractional exciton binding energy decreases linearly in line with the linear decrease of $E_{cont}$ (as seen in 
Fig. 1a).  In Fig. 3b we plot the excitonic binding energies for the $E_{11}$ excitons.  Given
that $K_{c+}^{-1}\sim 4$ for these gaps, we see that from Fig. 3a the binding energies roughly equal the gaps, $E_{11}$, themselves.
The estimates of the binding energies for the $E_{11}$ excitons of Ref. \cite{sfeir_nano} are considerably larger than those 
in Ref. \cite{louie1}, a consequence of our much larger estimate here of the renormalized band gaps.  It would thus be of considerable interest if these band
gaps could be measured directly.  But this is a difficult task as the standard method for measuring the particle-hole continuum,
scanning tunneling microscopy \cite{lin}, involves placing the tubes on a metallic substrate.  The consequent screening of the 
Coulomb interaction leads to values of $K_{c+}$ near to 1, far away from values of $K_{c+}$ appropriate for the excitons
measured in Ref. \cite{bnl_nano,sfeir_nano}.

In summary, we have presented a quantum field theoretical formalism able to predict
the absolute magnitudes of optically active excitons in 
semi-conducting carbon nanotubes over a wide range of diameters,
subbands, and dielectric environments.  This method involves a single
fitting parameter, $B$, relating the effective bandwidth of the tube
to the tube's diameter.  Once this parameter is in hand, a simple
scaling function yields the excitonic gaps for arbitrary nanotubes. We
have compared the predictions of this formalism with the excitonic data
of Refs. \cite{bnl_nano,sfeir_nano} and have found good agreement.

\section{Acknowledgements}

The research herein was 
carried out in part in the CMPMS Dept. (RMK and JM) and at
the Center for Functional Nanomaterials (MS), Brookhaven National Laboratory,
which is supported by the U.S. Department of Energy, Office of Basic
Energy Sciences, under Contract No. DE-AC02-98CH10886.

\break

\onecolumngrid

\section{Supplementary Material}

\subsection{Tables Comparing Measured Excitonic Gaps with Theoretical Predictions}

We report in Tables \ref{table1} and \ref{table2} the data for the excitons
measured in Ref. \cite{bnl_nano} and Ref. \cite{sfeir_nano}.

\begin{table*}[h]
\begin{tabular}{|l|l|l|l|l|l|l|l||l|l|l|l|l|l|l|}
\hline
(n,m) & $K_{c+},i=1,2$ & $\Delta_{0,11}$ & $E_{11,Th}$ & $E_{11,Exp}$ & $\Delta_{0,22}$ & $E_{22,Th}$ & $E_{22,Exp}$ & $K_{c+},i=3,4$ & $\Delta_{0,33}$ & $E_{33,Th}$ & $E_{33,Exp}$ & $\Delta_{0,44}$ & $E_{44,Th}$ & $E_{44,Exp}$ \\
\hline
(14,13) & 0.241 & 0.232 & 0.56 & 0.55 & 0.466 & 1.00  & 0.96 & 0.156 & 0.913 & 1.89 & 1.89 & 1.139 & 2.36 & 2.34 \\
\hline
(19,14) & 0.244 & 0.190 & 0.45 & 0.45 & 0.377 & 0.80 & 0.78 & 0.158 & 0.759 & 1.57  & 1.64 & 0.921 & 1.91 & 1.87 \\
\hline
(17,12) & 0.242 & 0.216 & 0.52 & 0.53 & 0.427 & 0.92 & 0.92 & 0.157 & 0.863 & 1.79 & 1.89 & 1.039 & 2.16 & 2.18 \\
\hline
(18,13) & 0.243 & 0.202 & 0.48 & 0.48 & 0.400 & 0.86 & 0.84 & 0.158 & 0.808 & 1.68 & 1.76 & 0.977 & 2.03 & 2.03 \\
\hline
\end{tabular}
\caption{\label{table1}
Comparison of the experimental and theoretical values of $E_{ii}$ of
the large radius tubes reported in Ref. \cite{bnl_nano}.  All energies
in units of eV.}
\vspace{-0.6cm}
\end{table*}

\begin{table*}[h]
\begin{tabular}{|l|l|l|l|l|}
\hline
(n,m) & $K_{c+}$ & $\Delta_{0,11}$ & $E_{11,Th}$ & $E_{11,Exp}$ \\
\hline
(8,3) & 0.26 & 0.562 & 1.35 & 1.30\\
\hline
(6,5) & 0.26 & 0.564 & 1.36 & 1.26\\
\hline
(7,5) & 0.26 & 0.523 & 1.25 & 1.21\\
\hline
(10,2) & 0.26 & 0.499 & 1.20 & 1.18\\
\hline
(9.4) & 0.26 & 0.478 & 1.15 & 1.13\\
\hline
(7,6) & 0.26 & 0.479 & 1.15 & 1.1\\
\hline
(8,6) & 0.26 & 0.448 & 1.08 & 1.06\\
\hline
(11,3) & 0.26 & 0.434 & 1.04 & 1.04\\
\hline
(9,5) & 0.26 & 0.436 & 1.03 & 1.00\\
\hline
(8,7) & 0.26 & 0.416 & 0.97 & 0.98\\
\hline
(9,7) & 0.26 & 0.392 & 0.94 & 0.94\\
\hline
(12,4) & 0.26 & 0.383 & 0.92 & 0.92\\
\hline
(11,6) & 0.26 & 0.367 & 0.88 & 0.89\\
\hline
\end{tabular}
\caption{\label{table2}
Comparison of the experimental and theoretical values of $E_{11}$ of
the small radius tubes reported in Ref. \cite{sfeir_nano}.}
\vspace{-0.6cm}
\end{table*}

\subsection{Derivation of the Scaling Function Governing the Excitonic Gaps}

In order to derive the form of the scaling function in Eqn. (\ref{scaling}), we need to first understand how a cutoff, 
which we will call $\Lambda_{TCSA}$, is implemented in the numerical methodology, the TCSA+NRG \cite{konik}, used to
study this system.  This method is able to study any Hamiltonian which can be written as a perturbed conformal field theory:
\begin{equation}
{\cal H} = {\cal H}_{CFT}+\lambda\Phi_{\rm perturbation},
\end{equation}
where here in this case ${\cal H}_{CFT}$ is a theory of four bosons, $\theta_i$, the coupling $\lambda$ equals $4\tilde\Delta_0/\pi$, and $\Phi_{\rm perturbation}=(\prod_{i=1}^4\cos(\theta_i/2)+\prod_{i=1}^4\sin(\theta_i/2))$.  The method uses the Hilbert space of ${\cal H}_{CFT}$ as a computational
basis.  (For details see Refs. \cite{konik,kon1}.)  This computational basis is optimal because the exact computation of matrix elements of $\Phi_{\rm perturbation}$ is readily
done using the commutation relations of the governing algebra of the unperturbed conformal theory, the Virasoro algebra.  Being
able to compute these matrix elements
means ${\cal H}$ can be recast as a matrix.  For this matrix to be a finite matrix, we need to truncate the Hilbert space of ${\cal H}_{CFT}$.
The unperturbed energies of the eigenstates of ${\cal H}_{CFT}$, $\{|\beta\rangle\}$, appearing in the excitonic sector have the form
\begin{equation}
E_\beta = \sum_{i=1}^4 \frac{v_0}{K_i}(\frac{2\pi n_i}{R} - \frac{c}{12}),
\end{equation}
where the $n_i$ are integers and $c$ is the central charge of a single boson ($c=1$).  To implement the cutoff we then insist that
the integers, $n_i$, satisfy $\sum_i (n_i/K_{i}) \leq N$.  This allows us to define the cutoff of this method as
\begin{equation}
\Lambda_{TCSA} = v_0\frac{2\pi N}{R}.
\end{equation}

With this in hand, the next step in the derivation of the scaling form is to write down the $\beta$-function of the coupling constant
$\tilde\Delta$:
\begin{equation}
N\frac{d\tilde\Delta_0}{dN} = \alpha(K_{c+}) \frac{\tilde\Delta_0^3}{v_0^2} \bigg(\frac{R}{2\pi N}\bigg)^{(5-K_{c+})/2}.
\end{equation}
In principle $\alpha(K_{c+})$ can be determined analytically -- we however extract it numerically from the TCSA data.
This numerical determination is what is used to plot $A(K_{c+})$ in Fig. 1b.
The form this $\beta$-function has can be determined following Ref. \cite{Watts} by insisting that the partition function of the theory remains
invariant under changes in the cutoff $N$.  In the gapped phase of the theory with $R$ sufficiently large this is equivalent to insisting
the gaps of the theory are invariant under the RG flow.  

If we integrate this $\beta$-function we obtain an expression relating the coupling in the absence of a cutoff to that with a cutoff:
\begin{equation}\label{intbfun}
\tilde\Delta_0(N=\infty) = \frac{\tilde\Delta_0(N)}{1-\frac{4\alpha (K_{c+})}{5-K_{c+}} \frac{\tilde\Delta_0^2(N)}{v_0^2} (\frac{R}{2\pi N})^{(5-K_{c+})/2}}.
\end{equation}
The gaps, $E_\alpha$, in the absence of a cutoff, depend on the coupling $\tilde\Delta_0(\infty)$ via the relation,
\begin{equation}\label{mass_rel}
E_\alpha (N=\infty,\tilde\Delta_0=\tilde\Delta_0(\infty)) = f^\infty_\alpha \tilde\Delta_0(\infty)^{4/(5-K_{c+})},
\end{equation}
a simple consequence of dimensional analysis (taking into account the anomalous dimensions of the coupling constant, $\tilde\Delta_0$).
By RG invariance we have $E_\alpha(N=\infty,\tilde\Delta(\infty))=E_\alpha(N,\tilde\Delta(N))$.  So substituting
this into Eqn. \ref{mass_rel} and using Eqn. \ref{intbfun} we obtain the desired scaling form:
\begin{eqnarray}\label{scaling1}
E_\alpha (N,\tilde\Delta_0(N)) &=& f^\infty_\alpha (\tilde\Delta_0(N))^{4/(5-K_{c+})}\times\cr\cr
&& \hskip -1in\bigg(1+ \frac{16\alpha(K_{c+})}{(5-K_{c+})^2} (\frac{\tilde\Delta_0}{v_0})^2(\frac{R}{2\pi N})^{(5-K_{c+})/2}\bigg).
\end{eqnarray}
The only issue is that this is expressed in terms of the TCSA cutoff $\Lambda_{TCSA}=2\pi N/R$ that arises from our
numerical treatment of the problem and not the
effective bandwidth of the tube, $\Lambda$.

To determine the relationship between $\Lambda$ and $\Lambda_{TCSA}$ we begin by consider the bosonization formula
giving the right/left moving fermion, $\psi^\dagger_\pm$, in terms of a normal ordered vertex operator of a boson:
\begin{equation}
\psi^\dagger_\pm (x) \sim :e^{i\phi_\pm (x)}\!:.
\end{equation}
In writing this expression we have dropped prefactors, zero modes, and Klein factors -- for our purposes what matters is
the normal ordered exponential.  The key to the relationship between $\Lambda$ and $\Lambda_{TCSA}$ is found
in the relation between the normal ordered vertex operator and its unnormal ordered counterpart:
\begin{eqnarray}
:e^{i\phi (x)}\!: &=& \sqrt{\frac{2\pi}{R}}e^{i\phi_\pm (x)} e^{\frac{1}{2}\sum^N_{n>0}\frac{1}{n}}\cr\cr
&\approx& \sqrt{\frac{2\pi}{L}} e^{i\phi_\pm (x)}e^{\gamma/2}N^{1/2},
\end{eqnarray}
where $\gamma$ is the Euler constant and the factor $\sqrt{2\pi/L}$ ensures the engineering dimension of the normal ordered vertex
operator matches its anomalous dimension.  The appearance of $N=R\Lambda_{TCSA}/2\pi$ reflects our use of the TCSA cutoff
to regulate the UV divergences that normal ordering exhibits in the theory.

When we initially bosonize the theory, the total charge boson is normal ordered assuming $K_{c+}=1$.  When we rediagonalize
the theory, absorbing the forward scattering part of the Coulomb interaction into the quadratic part of ${\cal H}$, we have to adjust the
normal ordering to take into account $K_{c+}\neq 1$.  We do so as follows:
\begin{eqnarray}
:e^{\theta_{c+}/2}\!:_{K_{c+}=1}&=& (\frac{2\pi}{R})^{1/4}e^{\gamma/4}N_{c}^{1/4}e^{i\theta_{c+}/2}\cr\cr
&=&\bigg( e^\gamma\frac{2\pi N_c}{R}\bigg)^{(1-K_{c+})/4}:e^{\theta_{c+}/2}\!:_{K_{c+}\neq1},\nonumber
\end{eqnarray}
where the subscripts $::_{K_{c+}}$ indicate the value of $K_{c+}$ for which the normal ordering is being done.  We use $N_c$
instead of $N$ as $N_c$ governs the maximal energy in the total charge (c+) sector of the theory, not the entire theory
itself.  The two are related via 
\begin{equation}
N_{c+} = \frac{Nv_0}{4v_{c+}},
\end{equation}
assuming an equipartition of energy between the four bosons in the theory.
It is this difference in normal ordering prefactors that is absorbed into the bare coupling:
\begin{equation}
\tilde\Delta_0 = \Delta_0 \bigg( e^\gamma\frac{2\pi N_{c+}}{R}\bigg)^{(1-K_{c+})/4}.
\end{equation}
This then implies (comparing the above with the relation below Eqn. (\ref{hgap}) in the main body of the text)
\begin{equation}
\frac{\Lambda}{v_{c+}} = \frac{e^\gamma}{4v_{c+}} \Lambda_{TCSA} = \frac{B}{d}.
\end{equation}
With this relation, we can now place the scaling function into its final form, Eqn. (\ref{scaling}), substituting $\Lambda$ for
$\Lambda_{TSCA}$ in Eqn. (\ref{scaling1}).

\subsection{Corrections to Excitonic Energies due to Intersubband Interactions}
In this section we will compute the corrections to excitonic energies due to interactions between subbands.  We will demonstrate that
they are proportional to $v_F^4/c^4$ where $c$ is the speed of light and so is small.

Consider an
excitonic excitation in subband i with energy $\Delta_i$.  The forward scattering portion of the intersubband
Coulomb interaction (as with the intrasubband interactions, the strongest part of the Coulomb interaction) takes the form
\begin{equation}
H_{\rm inter CI} = \sum_{i > j} \int dx dx'\rho_i(x)V_c(x-x')\rho_j(x')
\end{equation}
where $\rho_i$ is the density in the $i-$th subband.  In the long wavelength limit this can be rewritten as
\begin{eqnarray}
H_{\rm inter CI} &=& \sum_{i > j} \int dx \rho_i(x)\rho_j(x) V_c(k=0)\cr\cr 
&=& \Gamma\sum_{i >j} \int dx \partial_x \theta_{c+,i}(x)\partial_x\theta_{c+,j}(x),
\end{eqnarray}
where $\Gamma= v_F/(8\pi K_{c+})$ and we have used in the second line the bosonized expressions for the electron densities in the subbands.

In second order perturbation theory the correction to $\Delta_i$ takes the form 
\begin{equation}
\delta\Delta_i = \Gamma^2 \sum_n \frac{|\langle \Delta_i|\otimes\langle GS_j|H_{\rm inter CI}|GS_i\rangle\otimes |\Delta_{n,j}\rangle|^2}{\Delta_i-\Delta_{n,j}}
\end{equation}
where $|\Delta_{n,j}\rangle$ is some excitation in the $j-$th subband with parity odd symmetry (i.e. odd under $\theta_{c+,j}\rightarrow -\theta_{c+,j}$)
with energy $\Delta_{n,j}$.  The lowest energy such excitations are the one-photon excitons in subband $j$.  The state $|GS_j\rangle$ is
the ground state of the $j-$th subband.  The matrix elements that we to evaluate
in this sum take the form
\begin{eqnarray}
\langle \Delta_i|\rho_i (x)|GS_i\rangle &=& M_ie^{ip_ix}p_i;\cr\cr
\langle \Delta_{n,j}|\rho_j (x)|GS_j\rangle &=& M_{n,j}e^{ip_jx}p_{j,n},
\end{eqnarray}
where $M_i$ and $M_{n,j}$ are ${\cal O}(1)$ constants (as can be verified numerically) and $p_i/p_{n,j}$ are the momenta of
the excitations $|\Delta_i\rangle$/$|\Delta_{n,j}\rangle$.  
Thus the energy correction takes the form 
\begin{equation}
\delta\Delta_i = \sum_n \frac{v_F^2p_1^4|M_i|^2|M_{j,n}|^2\Gamma^2}{(\Delta_i^2+v_F^2p_i^2)(\Delta_i-E_{n,j})}.
\end{equation}
As one can see this correction vanishes as the momentum of the exciton goes to zero.  Typically the momentum of an optically
excited exciton will be equal to $\Delta_i/c$, implying that $\delta\Delta_i$ is proportional to $(v_F/c)^4$ and so is very small.

\end{document}